\documentclass[prd,preprint,superscriptaddress,showpacs,nofootinbib,%
tightenlines]{revtex4} 
\usepackage{graphics} 
\usepackage{epsfig}

\newcommand{\ben}{\begin{displaymath}} 
\newcommand{\een}{\end{displaymath}} 
\newcommand{\be}{\begin{equation}} 
\newcommand{\ee}{\end{equation}} 
\newcommand{\bea}{\begin{eqnarray}} 
\newcommand{\eea}{\end{eqnarray}} 
\begin{document}
\preprint{MKPH-T-03-2}
\title{Renormalization of relativistic baryon chiral perturbation theory 
and power counting} 
\author{T.~Fuchs} 
\affiliation{Institut f\"ur Kernphysik, Johannes 
Gutenberg-Universit\"at, D-55099 Mainz, Germany} 
\author{J.~Gegelia} 
\thanks{Alexander von Humboldt Research Fellow} 
\affiliation{Institut f\"ur Kernphysik, Johannes 
Gutenberg-Universit\"at, D-55099 Mainz, Germany} 
\affiliation{High Energy Physics Institute, 
Tbilisi State University, 
University St.~9, 380086 Tbilisi, Georgia} 
\author{G.~Japaridze}
\affiliation{Center for Theoretical Studies of Physical Systems, 
Clark Atlanta University, Atlanta, Georgia 30314, USA}
\author{S.~Scherer} 
\affiliation{Institut f\"ur Kernphysik, Johannes 
Gutenberg-Universit\"at, D-55099 Mainz, Germany} 
\date{July 1, 2003}
\begin{abstract} 
   We discuss a renormalization scheme for relativistic baryon
chiral perturbation theory which provides a simple and consistent power 
counting for renormalized diagrams.
   The method involves finite subtractions of dimensionally regularized 
diagrams beyond the standard $\overline{\rm MS}$ scheme of chiral 
perturbation theory to remove contributions violating the power counting.
   This is achieved by a suitable renormalization of the parameters
of the most general effective Lagrangian.
   In addition to its simplicity our method has the benefit that it can be
easily applied to multiloop diagrams.
   As an application we discuss the mass of the nucleon and compare the
result with the expression of the infrared regularization of Becher and 
Leutwyler.
\end{abstract} 
\maketitle
 
\section{\label{introduction}Introduction} 
   Starting from the pioneering work of Weinberg in 1979 
\cite{Weinberg:1979kz}, effective field theory has evolved into one of the 
most important theoretical tools for investigating strong-interaction 
processes in the low-energy regime. 
   The concept of spontaneous symmetry breakdown, leading to the appearence of 
massless Goldstone bosons with vanishing interactions in the zero-energy 
limit, was already well-known in the beginning of the 1960s
\cite{Nambu:xd,Nambu:tp,Goldstone:eq,Goldstone:es}. 
   Explicit symmetry breaking was taken into account in the framework of 
current algebra in combination with
the partially conserved axial-vector current (PCAC) 
hypothesis \cite{Gell-Mann:1964tf}
(for an overview see, e.g., \cite{Adler:1968,Treiman:1972,DeAlfaro:1973}).
   Already in the 1960s, Weinberg realized that the predictions derived 
from current algebra 
could be reproduced in the framework of the so-called phenomenological 
approximation (tree-level diagrams) of an effective Lagrangian 
\cite{Weinberg:1966fm}. 
   The key progress due to Weinberg's approach in 1979 was to 
systematically analyze {\em corrections} to the leading soft-pion results 
invoking a perturbative scheme not in terms of a 
coupling constant but rather in terms of external momenta and the pion mass
\cite{Weinberg:1979kz}. 
   Because of spontaneous symmetry breaking such an expansion is expected 
to work for momenta which are small compared to some intrinsic scale of the 
underlying theory.
   Since the starting point is a nonrenormalizable theory, infinities 
encountered in the calculation of loop diagrams need to be removed by 
a renormalization of the infinite number of free parameters of the most 
general effective Lagrangian. 
   However, as long as one includes all of the infinite number of interactions
allowed by symmetries, from the point of view of removing divergences  there is
no difference between the so-called nonrenormalizable theories and
renormalizable theories \cite{Weinberg:mt}.
  As will be discussed later in detail, the freedom of choosing a
renormalization scheme \cite{Collins:xc} can be advantageously
used to formulate a power counting for the perturbative calculation of 
renormalized diagrams.

   The ideas of Weinberg were further developed and comprehensively applied 
by Gasser and Leutwyler \cite{Gasser:1984yg,Gasser:1984gg} in terms of the 
generating functional of color-neutral quark bilinears which, at low 
energies, is dominated by the exchange and interaction of Goldstone bosons. 
   A correspondence between the loop expansion and the chiral expansion 
in terms of momenta and quark masses at a fixed ratio was set up. 
   Chiral perturbation theory (ChPT) in the mesonic sector has generated a 
host of successful applications up to and including the two-loop level
(for a recent review see, e.g., Ref.\ \cite{Scherer:2002tk}).
   The extension to processes involving one external nucleon was developed 
by Gasser, Sainio, and \v{S}varc \cite{Gasser:1988rb}.
   One of the findings in their approach was that higher-loop diagrams can 
contribute to terms as low as ${\cal O}(q^2)$, where $q$ generically
denotes a small expansion parameter such as, e.g., the pion mass.
   This ``mismatch'' between the chiral and the loop expansion has widely
been interpreted as the absence of a systematic power counting
in the relativistic formulation.
   Gasser, Sainio, and \v{S}varc pointed out that the appearance of 
another scale, namely, the mass of the nucleon (which does not 
vanish in the chiral limit) is one of the origins for the complications
in the baryonic sector.
   The heavy-baryon formulation of ChPT \cite{Jenkins:1990jv,Bernard:1992qa} 
provides a power counting scheme which is very similar to the mesonic
sector.  
   The basic idea consists in expressing the relativistic nucleon field 
in terms of a velocity-dependent field, thus dividing nucleon momenta 
into a large piece close to on-shell kinematics and a soft residual 
contribution. 
   Most of the calculations in the one-baryon sector have been performed 
in this framework (for an overview see, e.g., Ref.\ \cite{Bernard:1995dp})
which essentially corresponds to a simultaneous expansion 
of matrix elements in $1/m_N$ and $1/(4\pi F_\pi)$. 
   Although this scheme leads to a straightforward power counting, 
its disadvantage is that, in some cases, it does not provide the 
correct analytic behavior even in the threshold regime
\cite{Bernard:1996cc}. 
   Several methods have been suggested to reconcile power counting with the
constraints of analyticity in the relativistic approach 
\cite{Tang:1996ca,Ellis:1997kc,Becher:1999he,Gegelia:1999gf,Gegelia:1999qt,%
Lutz:1999yr,Lutz:2001yb}.
   The most widely used technique is the so-called infrared regularization
of Becher and Leutwyler \cite{Becher:1999he}
which has been applied in various calculations of baryon properties 
\cite{Ellis:1999jt,Kubis:2000zd,Zhu:2000zf,Kubis:2000aa,Zhu:2002tn},
pion-nucleon scattering \cite{Becher:2001hv,Torikoshi:2002bt},
mesonic U(3) chiral perturbation theory 
\cite{Borasoy:2001ik,Beisert:2001qb,Beisert:2002ad},
a discussion of the generalized Gerasimov-Drell-Hearn sum rule 
and the spin structure of the nucleon
\cite{Bernard:2002bs,Bernard:2002pw},
and the ground-state energy of pionic hydrogen \cite{Gasser:2002am}.

   The purpose of this work is to devise a new renormalization scheme
leading to a simple and consistent power counting for the renormalized diagrams
of a relativistic approach.
   The basic idea consists in performing additional subtractions of 
dimensionally regularized diagrams beyond the modified minimal subtraction 
scheme employed in Ref.\ \cite{Gasser:1988rb}.
   Our approach is motivated by an observation made in the context
of nonrelativistic nucleon-nucleon scattering, where the
application of the minimal subtraction scheme proved to be problematic.
   It was shown that the use of an appropriately chosen renormalization 
condition allows one to solve the problem of an ``unnaturally'' large 
scattering length and to obtain a consistent power counting in the 
two-nucleon sector 
\cite{Lepage:1997cs,Kaplan:1998tg,Gegelia:1998xr}. 
   Essentially the same idea of using a suitable renormalization 
condition has been discussed in Refs.\ \cite{Gegelia:1999gf,Gegelia:1999qt}
for a simplified model of the one-nucleon sector of
relativistic baryon chiral perturbation theory.
   One of the advantages of this approach, besides its simplicity, 
is that it may also be easily used in the renormalization of higher-order loop 
diagrams. 
 
  Our work is organized as follows. 
  In Sec.\ \ref{effective_Lagrangian} 
we provide those elements of the most general effective 
Lagrangian which are relevant for the calculation of the nucleon self-energy. 
  In Sec.\ \ref{eomsrvirbl} we illustrate our method by means of a simple 
dimensionally regularized one-loop integral and compare the result with the 
infrared regularization of  Becher and Leutwyler.
   In Sec.\ \ref{nucleon_self_energy} we apply our renormalization scheme to 
the calculation of the nucleon mass.
   General conclusions are presented in Sec.\ \ref{conclusions}.   

\section{The effective Lagrangian} 
\label{effective_Lagrangian}
   In this section we will briefly discuss those elements of the 
most general effective Lagrangian in the single-nucleon sector 
which are relevant for the subsequent calculation of the nucleon self-energy. 
   The effective Lagrangian consists of the sum of the purely mesonic 
and the $\pi N$ Lagrangians, respectively, 
\begin{displaymath} 
{\cal L}_{\rm eff}={\cal L}_{\pi}+{\cal L}_{\pi N}, 
\end{displaymath} 
both of which are organized in a (chiral) derivative and quark-mass expansion 
\cite{Weinberg:1979kz,Gasser:1984yg,Gasser:1984gg,Gasser:1988rb,%
Fearing:1994ga,Bijnens:1999sh,Fettes:2000gb,%
Ebertshauser:2001nj,Bijnens:2001bb}, 
\begin{eqnarray*} 
{\cal L}_{\pi}&=&{\cal L}_2+{\cal L}_4+{\cal L}_6+\cdots,\\ 
{\cal L}_{\pi N}&=&{\cal L}_{\pi N}^{(1)}+{\cal L}_{\pi N}^{(2)}+ 
{\cal L}_{\pi N}^{(3)} 
+{\cal L}_{\pi N}^{(4)}+\cdots, 
\end{eqnarray*} 
where the subscripts (superscripts) in ${\cal L}_{\pi}$ (${\cal L}_{\pi N}$) 
refer to the order in the expansion. 
   Counting the quark-mass term as ${\cal O}(q^2)$ 
\cite{Gasser:1984yg,Colangelo:2001sp}, 
the mesonic Lagrangian contains only even powers, whereas the baryonic 
Lagrangian involves both even and odd powers due to the additional spin 
degree of freedom. 
 
   From the mesonic sector we only need the 
lowest-order Lagrangian [${\cal O}(q^2)$] \cite{Gasser:1984yg}, 
\begin{equation} 
\label{l2} 
{\cal L}_2=\frac{F^2}{4}\mbox{Tr}(\partial_\mu U \partial^\mu U^\dagger) 
+\frac{F^2 M^2}{4}\mbox{Tr}(U^\dagger+ U), 
\end{equation} 
where $U$ is a unimodular unitary $(2\times 2)$ matrix containing 
the Goldstone boson fields. 
   In Eq.\ (\ref{l2}), $F$ denotes the pion-decay constant in the chiral 
limit: $F_\pi=F[1+{\cal O}(\hat{m})]=92.4$ MeV. 
   Here, we work in the isospin-symmetric limit $m_u=m_d=\hat{m}$, 
and the lowest-order expression for the squared pion mass is 
$M^2=2 B \hat{m}$, where $B$ is related to the quark condensate 
$\langle \bar{q} q\rangle_0$ in the chiral limit \cite{Gasser:1984yg}. 
 
   In order to discuss the $\pi N$ Lagrangian, let 
\begin{displaymath} 
\Psi=\left(\begin{array}{c}p\\n\end{array}\right) 
\end{displaymath} 
denote the nucleon field with two four-component Dirac fields $p$ and $n$ 
describing the proton and neutron, respectively. 
   The most general $\pi N$ Lagrangian ${\cal L}_{\pi N}$ is bilinear in 
$\bar\Psi(x)$ and $\Psi(x)$ and involves the quantities $u$, $u_\mu$, 
$\Gamma_\mu$, and $\chi_{\pm}$ (and their derivatives), 
which, in the absence of external fields, read 
$$ 
u^2=U,\quad 
u_\mu =iu^{\dagger}\partial_\mu U u^{\dagger},\quad 
\Gamma_\mu =\frac{1}{2} [u^{\dagger},\partial_\mu u], \quad 
\chi_{\pm}=M^2(U^{\dagger}\pm U). 
$$ 
   In terms of these building blocks the lowest-order Lagrangian reads 
\cite{Gasser:1988rb} 
\begin{equation} 
{\cal L}_{\pi N}^{(1)}=\bar \Psi \left( i\gamma_\mu D^\mu -m 
+\frac{1}{2} \stackrel{\circ}{g_A}\gamma_\mu \gamma_5 u^\mu\right) \Psi, 
\label{lolagr} 
\end{equation} 
   where $D_\mu\Psi = (\partial_\mu +\Gamma_\mu)\Psi $ denotes the 
covariant derivative (in the absence of external vector and axial-vector 
fields) and $m$ and $\stackrel{\circ}{g_A}$ refer to the chiral limit 
of the physical nucleon mass and the axial-vector coupling constant, 
respectively. 
   We have not displayed the corresponding counterterms in 
${\cal L}_{\pi N}^{(1)}$ which are understood to be fixed in such a manner 
that the pole position of the nucleon propagator as well as the axial-vector 
coupling constant (in the chiral limit) are not affected by loop 
contributions. 
   The explicit expressions of these counterterms in lowest order were 
identified in Ref.\ \cite{Gasser:1988rb}. 
 
   For our purposes, we only need to consider four of the seven structures 
of the Lagrangian at ${\cal O}(q^2)$ \cite{Gasser:1988rb,Fettes:2000gb}, 
\begin{eqnarray} 
{\cal L}_{\pi N}^{(2)}&=&c_1 \mbox{Tr}(\chi_{+})\bar\Psi\Psi 
-\frac{c_2}{4m^2}\mbox{Tr}(u_\mu u_\nu) 
\left(\bar\Psi D^\mu D^\nu\Psi+\mbox{H.c.} \right)\nonumber\\ 
&&+\frac{c_3}{2}\mbox{Tr}(u^\mu u_\mu)\bar\Psi\Psi 
-\frac{c_4}{4}\bar\Psi\gamma^\mu\gamma^\nu [u_\mu ,u_\nu ]\Psi+\cdots, 
\label{p2olagr} 
\end{eqnarray} 
where H.c.\ refers to the Hermitian conjugate. 
   While the Lagrangian ${\cal L}^{(3)}_{\pi N}$ does not contribute to the 
nucleon mass, at ${\cal O}(q^4)$ we need to consider the term
\begin{equation} 
\label{defalphabeta} 
-\frac{\alpha}{2}M^4 \bar{\Psi}\Psi,
\end{equation} 
resulting in the contribution $\alpha M^4/2$ to the nucleon mass. 
   This term results from identifying the relevant part of the most 
general chiral Lagrangian at ${\cal O}(q^4)$. 
   To be specific, the coefficient $\alpha$ of Eq.\ 
(\ref{defalphabeta}) is related to 
the parameters $e_i$ of the Lagrangian ${\cal L}_{\pi N}^{(4)}$ of Ref.\ 
\cite{Fettes:2000gb} by 
\begin{equation} 
\label{abci} 
\alpha=-4(8e_{38}+ e_{115}+e_{116}).
\end{equation} 
 
\section{\label{eomsrvirbl} 
Extended On-Mass-Shell Renormalization versus the infrared 
regularization of Becher and Leutwyler} 
   The basic idea of our renormalization scheme consists of providing   
a rule determining which terms of a given diagram should be subtracted
in order to satisfy a ``naive'' power counting by which one associates a 
well-defined power with the diagram in question.
   The terms to be subtracted are polynomials in small variables
and parameters (external momenta and squared pion mass) and can thus be 
realized by a suitable adjustment of the counterterms of the most general 
effective Lagrangian. 
   In other words, our proposition is to perform additional
subtractions (finite in number) of dimensionally regularized diagrams beyond 
the modified minimal subtraction scheme employed in Ref.\ \cite{Gasser:1988rb}.

   In order to illustrate our method and to compare it with the approach 
of Becher and Leutwyler \cite{Becher:1999he}, we will first consider as 
an example the dimensionally regularized one-loop integral 
\begin{equation} 
\label{hdef} 
H(p^2,m^2,M^2;n)\equiv -i\int \frac{d^n k}{(2\pi)^n} 
\frac{1}{[(p-k)^2-m^2+i0^+][k^2-M^2+i0^+]}, 
\end{equation} 
where $n$ denotes the number of space-time dimensions. 
   The masses $m$ and $M$ refer to the (lowest-order) nucleon and pion 
masses, respectively. 
   Such a type of integral is needed in, e.g., the calculation of the 
one-pion-loop contribution to the nucleon self-energy \cite{Gasser:1988rb}. 
   Dimensional regularization provides a convenient tool to handle the 
ultraviolet divergence resulting from the region where all components of 
$k^\mu$ get large. 
   However, as it stands, the loop integral of Eq.\ (\ref{hdef}) does not yet 
satisfy a simple chiral power counting. 
   Using Eq.\ (\ref{hdef}) we will propose a renormalization procedure 
generating a power counting for tree-level and loop diagrams of the 
{\em relativistic} effective field theory (REFT) which is analogous to that 
given in Ref.\ \cite{Weinberg:1991um} (for nonrelativistic nucleons). 
   As will be explained below, by subtracting a suitable number of 
counterterms in the integrand,\footnote{Here we make use of the fact that we 
may take more subtractions than would actually be necessary for the sole 
purpose of enforcing (ultraviolet) convergence.} 
we apply a renormalization scheme 
resulting in an effective cutoff $Q$ for loop integrals which is of 
the order of some small expansion parameters such as 
$Q^2\approx M^2$ and $Q m \approx|p^2-m^2|$ for the 
specific case of Eq.\ (\ref{hdef}). 
   Using the forest formula of Zimmermann \cite{Collins:xc,Zimmermann} 
allows one to systematically deal with any diagram. 
   The relevant subtractions can be implemented by adjusting the coefficients 
of the most general effective Lagrangian, i.e., the corresponding 
counterterms are local (polynomial) in momentum \cite{Collins:xc}, 
which implies that only a finite number of counterterms are needed for 
the subtraction of a specific diagram. 
   In general, this will then allow us to apply the following power counting: 
a loop integration in $n$ dimensions counts as $Q^n$, 
pion and fermion propagators count as $Q^{-2}$ and 
$Q^{-1}$, respectively, vertices derived from ${\cal L}_{2k}$ and 
${\cal L}_{\pi N}^{(k)}$ count as $Q^{2k}$ and $Q^k$, respectively.
   In total this yields for the power $D$ of a diagram in the 
one-nucleon sector the standard formula 
\cite{Weinberg:1991um,Ecker:1995gg}
\begin{equation}
\label{dimension}
D=n N_L - 2 I_\pi - I_N +\sum_{k=1}^\infty 2k N^\pi_{2k}
+\sum_{k=1}^\infty k N_k^N,
\end{equation}
where $N_L$ is the number of independent loop momenta, $I_\pi$ the number
of internal pion lines, $I_N$ the number of internal nucleon lines,
$N_{2k}^\pi$ the number of vertices originating from ${\cal L}_{2k}$,
and $N_k^N$ the number of vertices originating from ${\cal L}_{\pi N}^{(k)}$.
   In the language of chiral perturbation theory, $Q$ counts 
as a small momentum, i.e., as ${\cal O}(q)$, 
with the net result that Eq.\ (\ref{hdef}), after renormalization, is 
expected to be of order ${\cal O}(q^{n-3})$.

   Let us turn to the discussion of Eq.\ (\ref{hdef}). 
   We make use of the Feynman parametrization 
\begin{equation} 
{1\over ab}=\int_0^1 {dz\over [az+b(1-z)]^2} 
\label{schwfeyn} 
\end{equation} 
with $a=(p-k)^2-m^2+i0^+$ and $b=k^2-M^2+i0^+$, interchange the 
order of integrations, and perform the shift $k\to k+zp$ to obtain 
\begin{equation} 
\label{hpar} 
H(p^2,m^2,M^2;n)=-i\int_0^1 dz \int \frac{d^n k}{(2\pi)^n} 
\frac{1}{[k^2+p^2 z(1-z)-m^2z-M^2(1-z)+i0^+]^2}. 
\end{equation} 
   Making use of 
\begin{displaymath} 
\int \frac{d^n k}{(2\pi)^n} \frac{(k^2)^p}{(k^2-A)^q} 
=\frac{i(-)^{p-q}}{(4\pi)^\frac{n}{2}} 
\frac{\Gamma\left(p+\frac{n}{2}\right)\Gamma\left(q-p-\frac{n}{2}\right)}{ 
\Gamma\left(\frac{n}{2}\right)\Gamma(q)} 
A^{p+\frac{n}{2}-q}, 
\end{displaymath} 
with $p=0$ and $q=2$, we find 
\begin{equation} 
\label{HPmMn} 
H(p^2,m^2,M^2;n)=\frac{1}{(4\pi)^\frac{n}{2}} 
\Gamma\left(2-\frac{n}{2}\right) 
\int_0^1 dz [A(z)]^{\frac{n}{2}-2}, 
\end{equation} 
where 
\begin{displaymath} 
A(z)=-p^2(1-z)z+m^2(1-z)+M^2z-i0^+. 
\end{displaymath} 
 
\subsection{\label{chiral_limit}Chiral limit} 
   For the sake of simplicity, let us for the moment restrict ourselves to 
the (chiral) limit $M^2=0$ 
and introduce 
\begin{displaymath} 
C(z,\Delta)=z^2-\Delta\, z(1-z)-i0^+,\quad \Delta=\frac{p^2-m^2}{m^2}, 
\end{displaymath} 
so that we obtain 
\begin{equation} 
\label{Hcl} 
H(p^2,m^2,0;n)=\kappa(m;n)\int_0^1 dz [C(z,\Delta)]^{\frac{n}{2}-2}, 
\end{equation} 
where 
\begin{equation} 
\label{kappa} 
\kappa(m;n)=\frac{\Gamma\left(2-\frac{n}{2}\right)}{(4\pi)^\frac{n}{2}} 
m^{n-4}. 
\end{equation} 
   For the purpose of evaluating the integral of Eq.\ (\ref{Hcl}) 
we write\footnote{The boundary condition is properly taken into 
account by replacing $m^2\to m^2-i0^+$ in the final expression.} 
\begin{displaymath} 
\int_0^1 dz [C(z,\Delta)]^{\frac{n}{2}-2}= 
(-\Delta)^{\frac{n}{2}-2}\int_0^1 dz\, z^{\frac{n}{2}-2} 
\left(1-\frac{1+\Delta}{\Delta}z\right)^{\frac{n}{2}-2} 
\end{displaymath} 
and apply Eqs.\ 15.3.1 and 15.3.4 of Ref.\ \cite{Abramowitz} to obtain 
\begin{equation} 
\label{Hclresult} 
H(p^2,m^2,0;n)=\kappa(m;n)\frac{\Gamma\left(\frac{n}{2}-1\right)}{\Gamma 
\left(\frac{n}{2}\right)}F\left(1,2-\frac{n}{2};\frac{n}{2};\frac{p^2}{m^2} 
\right), 
\end{equation} 
where $F(a,b;c;z)$ is the hypergeometric function \cite{Abramowitz}. 
   In order to discuss the power counting properties of $H$ (in the chiral 
limit), we make use of Eq.\ 15.3.6 of Ref.\ \cite{Abramowitz} to rewrite 
Eq.\ (\ref{Hclresult}) as 
\begin{eqnarray} 
\label{j0dircalcrewr} 
H(p^2,m^2,0;n)&=&\frac{m^{n-4}}{(4\pi)^\frac{n}{2}}\left[ 
\frac{\Gamma\left(2-\frac{n}{2}\right)}{n-3} 
F\left(1,2-\frac{n}{2};4-n;-\Delta\right)\right.\nonumber\\ 
&&\left.+(-\Delta)^{n-3}\,\Gamma\left(\frac{n}{2}-1\right) 
\Gamma(3-n)F\left(\frac{n}{2}-1,n-2;n-2;-\Delta\right)\right]. 
\end{eqnarray} 
   Making use of 
\begin{equation} 
\label{fexpansion} 
F(a,b;c;z)=1+\frac{ab}{c}z+\frac{a(a+1)b(b+1)}{c(c+1)}\frac{z^2}{2}+\cdots 
\end{equation} 
for $|z|<1$ and the fact that $\Delta$ counts as a small quantity of order 
${\cal O}(q)$, we immediately see that the first term of 
Eq.\ (\ref{j0dircalcrewr}) contains a contribution which does not 
satisfy the above power counting, i.e., which is not proportional 
to ${\cal O}(q)$ as $n\to 4$. 
   Using the expansion of Eq.\ (\ref{fexpansion}) together with 
$\Gamma(1+x)=x\Gamma(x)$ we obtain, as $n\to 4$, 
\begin{equation} 
H=\frac{m^{n-4}}{(4\pi )^{\frac{n}{2}}} 
\left[\frac{\Gamma \left(2-\frac{n}{2}\right)}{n-3} 
+\left( 1-{p^2\over m^2}\right)\ln \left( 1-{p^2\over m^2} 
\right)+ 
\left( 1-{p^2\over m^2}\right)^2\ln \left( 1-{p^2\over m^2}\right) 
+\cdots\right], 
\label{j0dircalcisolated} 
\end{equation} 
where $\cdots$ refers to terms which are at least of order ${\cal O}(q^3)$ 
or $O(n-4)$.\footnote{Note that we count a term of the type 
$-\Delta \ln(-\Delta)$ as ${\cal O}(q)$.} 
   If we subtract 
\begin{equation} 
\label{subtraction} 
\frac{m^{n-4}}{(4\pi )^{\frac{n}{2}}} 
\frac{\Gamma \left( 2-\frac{n}{2}\right)}{n-3} 
\label{subtrterm} 
\end{equation} 
from Eq.\ (\ref{j0dircalcisolated}) we obtain as the renormalized integral 
\begin{equation} 
H_R(p^2,m^2,0;n)={m^{n-4}\over (4\pi )^{n/2}}\left[ \left( 1-{p^2\over m^2} 
\right) 
\ln \left( 1-{p^2\over m^2}\right)+ 
\left( 1-{p^2\over m^2}\right)^2\ln \left( 1-{p^2\over m^2}\right) 
+\cdots\right]. 
\label{j0ren} 
\end{equation} 
   The subtracted term of Eq.\ (\ref{subtrterm}) is local in the external 
momentum $p$, i.e., it is a {\em polynomial} in $p^2$ 
and can thus be obtained by a {\em finite} number of counterterms in the 
most general effective Lagrangian. 
   In other words, using an ordinary subtractive renormalization with an 
appropriately chosen renormalization condition we obtained the renormalized 
expression of Eq.\ (\ref{j0ren}) which satisfies the power counting discussed 
above. 
 
   Using the example of Eq.\ (\ref{hdef}) (in the chiral limit) we now 
apply a conventional renormalization prescription which allows us to identify 
those terms which we subtract from a given integral without {\em explicitly} 
calculating the integral beforehand. 
   In essence we work with a modified integrand which is obtained from 
the original integrand by subtracting a suitable number of 
counterterms.\footnote{In the often used zero-momentum subtraction scheme 
a Taylor series expansion of the integrand with respect to the external 
momentum $p^\mu$ around $p^\mu=0$ is used.} 
   The meaning of suitable in the present context will be explained in 
a moment. 
   To that end we consider the series 
\begin{eqnarray} 
\label{series} 
&& 
\sum_{l=0}^\infty \frac{(p^2-m^2)^l}{l!}\left[ 
\left(\frac{1}{2p^2}p_\mu\frac{\partial}{\partial p_\mu}\right)^l 
\frac{1}{(k^2+i0^+)[k^2-2k\cdot p+(p^2-m^2)+i0^+]}\right]_{p^2=m^2}\nonumber\\ 
&=&\left.\frac{1}{(k^2+i0^+)(k^2-2k\cdot p+i0^+)}\right|_{p^2=m^2}\nonumber\\ 
&&+(p^2-m^2)\left[\frac{1}{2m^2}\frac{1}{(k^2-2k\cdot p+i0^+)^2} 
-\frac{1}{2m^2}\frac{1}{(k^2+i0^+)(k^2-2k\cdot p+i0^+)}\right.\nonumber\\ 
&&\left. -\frac{1}{(k^2+i0^+)(k^2-2k\cdot p+i0^+)^2} 
\right]_{p^2=m^2}\nonumber\\ 
&&+\cdots, 
\end{eqnarray} 
   where $[\cdots ]_{p^2=m^2}$ means that we consider the {\em coefficients} 
of $(p^2-m^2)^l$ only for four-momenta $p^\mu$ satisfying the on-mass-shell 
condition.\footnote{Equation (\ref{series}) is {\em not} a Taylor series 
of the integrand.} 
   Although the coefficients still depend on the direction of $p^\mu$, 
after integration of this series with respect to the loop momentum $k$ and 
evaluation of the resulting coefficients for $p^2=m^2$, the integrated series 
is a function of $p^2$ only. 
   In fact, as was shown in Ref.\ \cite{Gegelia:1994zz}, the integrated 
series exactly reproduces the first term of  Eq.\ (\ref{j0dircalcrewr}). 
   At this point we stress that 
\begin{displaymath} 
\left. 
-i\int \frac{d^n k}{(2\pi)^n} 
\frac{1}{(k^2+i0^+)(k^2-2k\cdot p+i0^+)}\right|_{p^2=m^2} 
\end{displaymath} 
and 
\begin{displaymath} 
\left[-i\int \frac{d^n k}{(2\pi)^n} 
\frac{1}{(k^2+i0^+)(k^2-2k\cdot p+p^2-m^2+ i0^+)}\right]_{p^2=m^2} 
\end{displaymath} 
are not the same for $n\leq 3$. 
   Let us provide a formal definition of our renormalization scheme: 
we subtract from the integrand of $H(p^2,m^2,0;n)$ those terms of the 
series of Eq.\ (\ref{series}) which violate the power counting. 
    These terms are always analytic in the small parameter and do not 
contain infrared singularities. 
   In the above example we only need to subtract the first term. 
   All the higher-order terms contain infrared singularities. 
   For example, the last term of the second coefficient would generate a 
behavior $k^3/k^4$ of the integrand for $n=4$. 
   The integral of the first term of Eq.\ (\ref{series}) is given 
by Eq.\ (\ref{subtraction}), and we end up with Eq.\ (\ref{j0ren}) for 
the renormalized integral. 
   Since we make use of the subtraction point $p^2=m^2$, we denote 
our renormalization condition ``extended on-mass-shell'' (EOMS) scheme 
in analogy with the on-mass-shell renormalization scheme in renormalizable 
theories. 
 
   Let us now compare with the approach of Becher and Leutwyler of Ref.\ 
\cite{Becher:1999he}, where the integral $H$ is divided into the 
so-called infrared (singular) part $I$ and the remainder $R$, 
defined as 
\begin{eqnarray} 
\label{idef} 
I&\equiv&\kappa(m;n)\int_0^\infty dz [C(z,\Delta)]^{\frac{n}{2}-2}, \\ 
\label{rdef} 
R&\equiv&-\kappa(m;n)\int_1^\infty dz [C(z,\Delta)]^{\frac{n}{2}-2}. 
\end{eqnarray} 
   The analytical expressions for both integrals are 
given by\footnote{The correct imaginary parts are obtained by replacing 
$m^2\to m^2-i0^+$.} 
\begin{eqnarray} 
\label{ianalytic} 
I&=&\frac{m^{n-4}}{(4\pi)^\frac{n}{2}}(-\Delta)^{n-3}\Gamma 
\left(\frac{n}{2}-1\right)\Gamma\left( 3-n\right) 
\frac{1}{(1+\Delta )^{\frac{n}{2}-1}},\\ 
\label{ranalytic} 
R&=&-\kappa(m;n)(1+\Delta)^{\frac{n}{2}-2} \frac{\Gamma(3-n)}{ 
\Gamma(4-n)}F\left(2-\frac{n}{2},3-n;4-n;\frac{\Delta}{1+\Delta}\right). 
\end{eqnarray} 
   Let us discuss a few properties of $I$ and $R$, respectively. 
   Counting $\Delta$ as a small quantity of ${\cal O}(q)$, the 
infrared part $I$ respects a simple power counting by being 
proportional to $q^{n-3}$. 
   As $n\to 4$, $I$ cannot be expanded in a power 
series in $\Delta$, because 
\begin{displaymath} 
(-\Delta)^{n-3}\Gamma(3-n)=-\Delta\left[\frac{1}{n-4}-\Gamma'(1)-1\right]
-\Delta \ln(-\Delta)+ O(n-4). 
\end{displaymath} 
   Finally, for noninteger values of $n$, $I$ contains noninteger powers of 
$\Delta$. 
   On the other hand, due to the analytic properties of the hypergeometric 
function, the remainder $R$ can be expanded in an ordinary 
Taylor series in $\Delta$ even for noninteger values of $n$. 
   However, as $\Delta\to 0$, $R$ does not fit into the above power counting, 
i.e., it is not proportional to a small quantity of order $q$ raised to the 
power $n-3$. 
   In the approach of Becher and Leutwyler one {\em explicitly} keeps the 
contribution $I$ of $H$ (with subtracted singularities when $n$ approaches 4) 
as the result of the integral and drops $R$ arguing that it 
is effectively taken into account through an infinite number of counterterms 
in the most general effective Lagrangian. 
   As pointed out in Ref.\ \cite{Becher:1999he}, the infrared part $I$ also 
contains an infinite number of divergent terms if expanded in powers of 
$\Delta$. 
    An infinite number of divergent terms in $R$ and $I$ exactly cancel each 
other and one is left with one ultraviolet divergent term in $H$ which is 
$\Delta$ independent, namely Eq.\ (\ref{subtraction}). 
 
\subsection{\label{finite_pion_mass}Finite pion mass} 
   We now generalize our renormalization scheme to the case of a nonvanishing 
pion mass [see Eq.\ (\ref{hdef})]. 
   For easier comparison with Ref.\ \cite{Becher:1999he} we introduce the 
variables 
\begin{equation} 
\label{Omegaalpha} 
\Omega ={p^2-m^2-M^2\over 2m M}, \quad 
\alpha ={M\over m}, 
\end{equation} 
where $\Omega$ counts as ${\cal O}(q^0)$ for $p^2\neq m^2$ 
[${\cal O}(q)$ for $p^2=m^2$] and $\alpha$ counts as ${\cal O}(q)$. 
   We obtain for Eq.\ (\ref{HPmMn}), as $n\to 4$,
\begin{equation} 
\label{Hresult} 
H=-2\bar\lambda +{1\over 16\pi^2}-{1\over 
8\pi^2}{\alpha\sqrt{1-\Omega^2}\over 
1+2\alpha\Omega +\alpha^2 }\arccos\left( -\Omega\right) 
-{1\over 8\pi^2}{\alpha (\alpha +\Omega)\over 1+2\alpha\Omega 
+\alpha^2} 
\ln(\alpha), 
\end{equation} 
where 
\begin{equation} 
\label{lambdabar} 
\bar\lambda ={m^{n-4}\over (4\pi )^2}\left\{ {1\over n-4}- 
{1\over 2}\left[ \ln (4\pi )+\Gamma '(1)+1\right]\right\}. 
\end{equation} 
   Clearly, the first two terms of Eq.\ (\ref{Hresult}) violate our power 
counting, since we want the renormalized integral to be of ${\cal O}(q)$ 
as $n\to 4$. 
 
   In order to apply our renormalization scheme to Eq.\ (\ref{hdef}), 
we observe that the dimensionally regularized integral contains a part 
which, for noninteger $n$, is proportional to noninteger powers of $M$ 
but does {\em not} violate the power counting. 
   On the other hand, the remaining piece of the integral may {\em always}, 
i.e., for arbitrary $n$, be expanded in non-negative powers of $M$, 
and it is this contribution which is responsible for the violation of power 
counting. 
   We expand this second part in terms of $M$ and $p^2-m^2$ and subtract those 
terms which violate the power counting. 
   In practice, we realize this scheme by writing down a series similar to 
Eq.\ (\ref{series}), where, in addition, we expand pion propagators 
in powers of $M^2$. 
   In the present case we only need to subtract the first term to satisfy 
the power counting: 
\begin{equation} 
H_{\rm subtr}=-i  \int {d^n k\over (2\pi )^n} 
\frac{1}{k^2 +i0^+}\left. 
\frac{1}{k^2-2p\cdot k +i0^+}\right|_{p^2=m^2}= 
-2\bar\lambda +{1\over 16\pi^2}+O(n-4). 
\label{simpleexnzmpexp} 
\end{equation} 
   Subtracting Eq.\ (\ref{simpleexnzmpexp}) from Eq.\ (\ref{Hresult}) our 
final expression for the renormalized integral reads 
\begin{equation} 
H_R=-{1\over 
8\pi^2}{\alpha\sqrt{1-\Omega^2}\over 
1+2\alpha\Omega +\alpha^2 }\arccos\left( -\Omega\right) 
-{1\over 8\pi^2}{\alpha (\alpha +\Omega)\over 1 
+2\alpha\Omega +\alpha^2} 
\ln(\alpha). 
\label{hren} 
\end{equation} 
   Again, the subtraction term $H_{\rm subtr}$ of Eq.\ (\ref{simpleexnzmpexp}) 
is local in the external momenta and can thus be realized as a counterterm 
in the most general effective Lagrangian. 
   Let us stress one more time that we count a term $\alpha\ln(\alpha)$ as 
${\cal O}(q)$. 
   Moreover, when expanded in small quantities, $H_R$ consists of an 
infinite string of terms of ${\cal O}(q^l)$ with $l\geq 1$. 
   In other words, when we say that an expression is of ${\cal O}(q)$, 
we refer to the {\em minimal} power $q^1$ of that expression. 
   This situation has to be contrasted with the mesonic sector, where 
an expression of, say, ${\cal O}(q^4)$ exclusively consists of terms 
of ${\cal O}(q^4)$. 
   We conclude that using ordinary renormalization with appropriately 
chosen renormalization conditions allows us to obtain the 
renormalized expression of Eq.\ (\ref{hren}) which satisfies power counting.

   It is instructive to compare Eq.\ (\ref{hren}) with the result of 
the infrared regularization of Becher and Leutwyler, where 
the integral $H$ is divided into an infrared part 
$I$ and a remainder $R$ \cite{Becher:1999he}: 
\begin{eqnarray} 
\label{ibl} 
I&=&\bar\lambda\nu -{1\over 8\pi^2}{\alpha\sqrt{1-\Omega^2}\over 
1+2\alpha\Omega +\alpha^2 }\arccos\left( -{\alpha +\Omega \over 
\sqrt{1+2\alpha\Omega +\alpha^2 }}\right)\nonumber\\ 
&&-{1\over 16\pi^2}{\alpha(\alpha +\Omega)\over 1+2\alpha\Omega 
+\alpha^2} 
[2\ln(\alpha) -1],\\ 
\label{rbl} 
R&=&-(2+\nu )\bar\lambda +{1\over 8\pi^2}{\alpha\sqrt{1-\Omega^2} 
\over 
1+2\alpha\Omega +\alpha^2 }\arcsin\left( {\alpha\sqrt{1-\Omega^2} 
\over \sqrt{1+2\alpha\Omega +\alpha^2 }}\right) 
+{1\over 16\pi^2}{1+\alpha\Omega\over 1+2\alpha\Omega +\alpha^2 }, 
\end{eqnarray} 
where 
\begin{displaymath} 
\nu =-{p^2-m^2+M^2\over p^2}. 
\end{displaymath} 
   Using elementary relations among the inverse trigonometric functions,
 the sum of $I$ and $R$ is indeed identical to 
Eq.\ (\ref{Hresult}). 
   In this decomposition $I$ satisfies the power counting whereas $R$, 
violating the power counting, is absorbed into an {\em infinite} number of 
counterterms. 
   The first (infinite) term of $I$ is also taken care of by renormalization. 
 
\section{\label{nucleon_self_energy}Nucleon self-energy} 
   As a specific example, we will now turn to the calculation of the nucleon 
self-energy at ${\cal O}(q^4)$. 
   The complete propagator of the nucleon is defined as the Fourier transform 
\begin{equation} 
\label{s0p} 
S_0(p)=\int d^4 x e^{ip\cdot x} S_0(x) 
\end{equation} 
of the two-point function 
\begin{equation} 
\label{s0x} 
S_0(x)=-i\langle 0|T[\Psi_0(x)\bar{\Psi}_0(0)]|0\rangle, 
\end{equation} 
where $\Psi_0$ denotes the bare nucleon field. 
   We parametrize 
\begin{equation} 
S_0(p)=\frac{1}{p\hspace{-.45em}/\hspace{.1em}-m_0 
-\Sigma_0(p\hspace{-.45em}/\hspace{.1em})}\equiv 
\frac{1}{p\hspace{-.45em}/\hspace{.1em}-m 
-\Sigma(p\hspace{-.45em}/\hspace{.1em})}, 
\end{equation} 
where $m_0$ refers to the bare mass of Eq.\ (\ref{pieceoflolagr}), 
whereas $m$ is the nucleon pole mass in the chiral limit. 
   Here, $\Sigma_0(p\hspace{-.45em}/\hspace{.1em})$ 
and $\Sigma(p\hspace{-.45em}/\hspace{.1em})$ 
are matrix functions \cite{Itzykson:rh} 
which, using 
$p\hspace{-.45em}/\hspace{.1em}p\hspace{-.45em}/\hspace{.1em}=p^2$, 
can be parametrized as 
\begin{displaymath} 
\Sigma_0(x)=-x f_0(x^2)+m_0 g_0(x^2) 
\end{displaymath} 
with an analogous expression for $\Sigma$. 
 
   We will express the nucleon self-energy 
$\Sigma(p\hspace{-.45em}/\hspace{.1em})$ in terms of $m$, 
the lowest-order pion mass $M$, and bare coupling constants. 
   In terms of Feynman diagrams, $-i\Sigma(p\hspace{-.45em}/\hspace{.1em})$ 
represents the one-particle-irreducible perturbative contribution to the 
two-point function. 
   Moreover, it also contains contributions of counterterms generated by 
$m_0$, which make sure that the pole mass in the chiral limit, $m$, 
stays put. 
   However, for the sake of simplicity we will not explicitly show 
these counterterms. 
 
   As usual the physical nucleon mass is defined through the pole 
of the full propagator at $p\hspace{-.45em}/\hspace{.1em}=m_N$, 
\begin{equation} 
\label{nucleonmassdefinition} 
m_N-m_0-\Sigma_0(m_N)=m_N-m-\Sigma(m_N)=0, 
\end{equation} 
while the wave function renormalization constant $Z_0$ is defined 
as the residue at $p\hspace{-.45em}/\hspace{.1em}=m_N$, 
\begin{eqnarray*} 
S_0(p)&=&\frac{1}{p\hspace{-.45em}/\hspace{.1em}-m_0-\Sigma_0(m_N 
+p\hspace{-.45em}/\hspace{.1em}-m_N)}\\ 
&=&\frac{1}{p\hspace{-.45em}/\hspace{.1em}-m_0-\Sigma_0(m_N) 
-(p\hspace{-.45em}/\hspace{.1em}-m_N)\Sigma'_0(m_N) 
+O[(p\hspace{-.45em}/\hspace{.1em}-m_N)^2]}\\ 
&=&\frac{1}{(p\hspace{-.45em}/\hspace{.1em}-m_N)[1-\Sigma'_0(m_N) 
+O(p\hspace{-.45em}/\hspace{.1em}-m_N)]}\\ 
&\to& \frac{Z_0}{p\hspace{-.45em}/\hspace{.1em}-m_N+i0^+}\,\, 
\mbox{for}\,\,p\hspace{-.45em}/\hspace{.1em}\to m_N, 
\end{eqnarray*} 
yielding 
\begin{equation} 
\label{z0def} 
Z_0=\frac{1}{1-\Sigma'_0(m_N)}=\frac{1}{1-\Sigma'(m_N)}. 
\end{equation} 
 
\begin{figure}
\epsfig{file=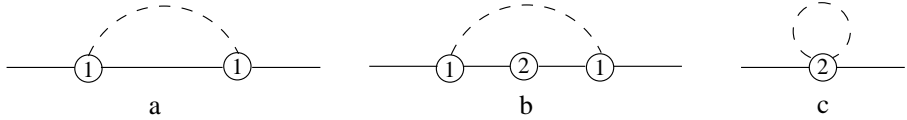, width=12truecm} 
\caption[]{\label{mnloops:fig}
One-loop contributions to the nucleon self-energy at ${\cal O}(q^4)$.
The numbers in the interaction blobs denote the order of the Lagrangian
from which they are obtained.}
\end{figure}

   At ${\cal O}(q^4)$, the self-energy receives contact contributions from 
${\cal L}_{\pi N}^{(2)}$ and ${\cal L}_{\pi N}^{(4)}$ as well as the 
one-loop contributions of Fig.\ \ref{mnloops:fig}, 
\begin{equation} 
\Sigma=\Sigma_{\rm contact}+\Sigma_{\rm loop}. 
\end{equation} 
   The contact contributions read 
\begin{equation} 
\label{sigmacontact} 
\Sigma_{\rm contact}=-4 M^2 c_1^0-2 M^4 (8e_{38}^0+e_{115}^0+e_{116}^0), 
\end{equation} 
where the superscripts 0 refer to bare parameters. 
   Applying Feynman rules we obtain three one-loop contributions (see Fig.\ 
\ref{mnloops:fig}) 
\begin{equation} 
\Sigma_{\rm loop} =\Sigma_a+\Sigma_b+\Sigma_c, 
\label{olse} 
\end{equation} 
where 
\begin{eqnarray} 
\label{sigmaa} \Sigma_a&=&\frac{3 {\stackrel{\circ}{g_{A}}}_0^2}{4 
F_0^2}\,i\int\frac{d^n k}{(2\pi)^n}\, 
k\hspace{-.45em}/\hspace{.1em}\gamma_5 \frac{1}{p\hspace{-.45em}/ 
\hspace{.1em}-k\hspace{-.45em}/\hspace{.1em}-m+i0^+}\, 
k\hspace{-.45em}/\hspace{.1em}\gamma_5\, 
\frac{1}{k^2-M^2+i0^+}\nonumber\\ 
&=&\frac{3 {\stackrel{\circ}{g_{A}}}_0^2}{4 F_0^2}
\,i\int\frac{d^n k}{(2\pi)^n}\, 
\frac{k\hspace{-.45em}/\hspace{.1em}(p\hspace{-.45em}/\hspace{.1em} 
-k\hspace{-.45em}/\hspace{.1em}-m)k\hspace{-.45em}/\hspace{.1em}}{
(k-p)^2-m^2+i0^+}\, \frac{1}{k^2-M^2+i0^+},\\ 
\label{sigmab} 
\Sigma_b&=&-4M^2 c_1^0\frac{3 {\stackrel{\circ}{g_{A}}}_0^2}{4 
F_0^2}\,i \int\frac{d^n k}{(2\pi)^n}\, 
k\hspace{-.45em}/\hspace{.1em}\gamma_5 
\left(\frac{1}{p\hspace{-.45em}/ 
\hspace{.1em}-k\hspace{-.45em}/\hspace{.1em}-m+i0^+}\right)^2\, 
k\hspace{-.45em}/\hspace{.1em}\gamma_5\,\frac{1}{k^2-M^2+i0^+}\nonumber\\ 
&=&-4M^2 c_1^0\,\frac{\partial\Sigma_a}{\partial m},\\ 
\label{sigmac} 
\Sigma_c&=& 3\frac{M^2}{F_0^2}\left(2 
c_1^0-c_3^0 - \frac{p^2}{m^2}\frac{c_2^0}{n}\right)i 
\int\frac{d^n k}{(2\pi)^n} \frac{1}{k^2-M^2+i0^+}. 
\end{eqnarray} 
   Using $\{\gamma^\mu,\gamma^\nu\}=2 g^{\mu\nu}$, Eq.\ (\ref{sigmaa}) 
can be expressed in terms of the basis integrals of Appendix 
\ref{loop_integrals} as 
\begin{eqnarray} 
\label{sigmaaresult} 
\Sigma_a&=&-\frac{3 {\stackrel{\circ}{g_{A}}}_0^2}{4 F_0^2}\left\{ 
\vphantom{-\frac{(p^2-m^2)p\hspace{-.5em}/}{2p^2}} 
(p\hspace{-.45em}/\hspace{.1em}+m)I_N 
+M^2(p\hspace{-.45em}/\hspace{.1em}+m)I_{N\pi}(-p,0)\right.\nonumber\\ 
&&\left. -\frac{(p^2-m^2)p\hspace{-.45em}/\hspace{.1em}}{2p^2}[ 
(p^2-m^2+M^2)I_{N\pi}(-p,0)+I_N-I_\pi]\right\}. 
\end{eqnarray} 
   The renormalization of the loop diagrams is performed in two steps. 
   First we render the diagrams finite by applying the subtraction scheme 
used by Gasser and Leutwyler \cite{Gasser:1984yg,Gasser:1988rb} 
which we denote by the modified minimal subtraction scheme of ChPT 
($\widetilde{\rm MS}$).\footnote{In distinction to the 
$\overline{\rm MS}$ scheme commonly used in Standard Model calculations, 
the $\widetilde{\rm MS}$ scheme  contains an additional finite subtraction 
term. 
   To be specific, in $\widetilde {\rm MS}$ one uses multiples 
of $ 1/(n-4)- \left[ \ln (4\pi )+\Gamma '(1)+1\right]/2$ 
instead of 
$1/(n-4)-\left[ \ln (4\pi )+\Gamma '(1)\right]/2$ 
in $\overline{\rm MS}$. 
} 
   We choose the renormalization parameter  (unit of mass 
or 't Hooft parameter) $\mu =m$. 
   In a second step we then perform additional {\em finite} subtractions for 
integrals which contain nucleon propagators with the purpose of imposing 
our power counting scheme. 
   In fact, in order to apply the $\widetilde{\rm MS}$ subtraction in 
practical calculations, we do not actually need to explicitly write down the 
corresponding counterterms. 
   We simply subtract all loop diagrams and replace the bare couplings with 
the couplings corresponding to the $\widetilde{\rm MS}$ scheme. 
   In the above expressions we replace subscripts and superscripts ``0'' 
denoting bare coupling constants with ``r'' and supply the integrals with 
indicators ``r'' referring to the fact that they have been subtracted. 
   For example, the result for $\Sigma_{r,a}$ then reads 
\begin{eqnarray} 
\label{sigmaaresultmsbar} 
\Sigma_{r,a}&=&-\frac{3 {\stackrel{\circ}{g_{A}}}_r^2}{4 F_r^2}\left\{ 
\vphantom{-\frac{(p^2-m^2)p\hspace{-.5em}/}{2p^2}} 
M^2(p\hspace{-.5em}/\hspace{.1em}+m)I_{N\pi}^r(-p,0)\right.\nonumber\\ 
&&\left. -\frac{(p^2-m^2)p\hspace{-.5em}/}{2p^2}[ 
(p^2-m^2+M^2)I_{N\pi}^r(-p,0)-I_\pi^r]\right\}, 
\end{eqnarray} 
   where the expressions for $I_\pi^r$ and $I^r_{N\pi}$ are given in Eqs.\
(\ref{Ipir}) and (\ref{INpir}) of Appendix \ref{loop_integrals}.
 
   The $\widetilde{\rm MS}$-subtracted self-energy corresponds to the Green 
function 
\begin{equation} 
\label{s1p} 
S_1(p)=\int d^4 x e^{ip\cdot x} S_1(x), 
\end{equation} 
where 
\begin{displaymath} 
S_1(x)=-i\langle 0|T[\Psi_1(x)\bar{\Psi}_1(0)]|0\rangle 
\end{displaymath} 
is the two-point function of the $\widetilde{\rm MS}$-renormalized field 
\begin{equation} 
\label{psi1} 
\Psi_1(x)\equiv\Psi_{\widetilde{\rm MS}}(x)=\frac{\Psi_0(x)}{\sqrt{Z_{01}}}. 
\end{equation} 
   We refer to $\sqrt{Z_{01}}$ as the field redefinition constant 
[see Eq.\ (\ref{renf})] connecting the bare field $\Psi_0$ and the 
$\widetilde{\rm MS}$-renormalized field $\Psi_1$. 
   Analogous to Eq.\ (\ref{nucleonmassdefinition}), the physical nucleon 
mass is determined through the pole of the $\widetilde{\rm MS}$-renormalized 
propagator. 
   We obtain for the mass in the $\widetilde{\rm MS}$ scheme
\begin{eqnarray} 
m_N&=&m-4 c_1^r M^2+\frac{3 {\stackrel{\circ}{g_{A}}}_r^2 M^2}{32\pi^2 F_r^2}m 
\left(1+8 c_1^r m\right) 
-\frac{3 {\stackrel{\circ}{g_{A}}}_r^2 M^3}{32 \pi F_r^2} 
\nonumber\\ 
&& 
+\frac{3  M^4}{32 \pi^2 F_r^2 }\ln\left(\frac{M}{m}\right) 
\left(8c_1^r-c_2^r-4 c_3^r-\frac{{\stackrel{\circ}{g_{A}}}_r^2}{m}\right) 
\nonumber\\ 
&& 
+\frac{3{\stackrel{\circ}{g_A}}_r^2 M^4}{32 \pi^2 F_r^2 m}
\left[1+4 c_1^r m\right] 
\nonumber\\ 
&& 
+M^4\left(\frac{3}{128\pi^2F_r^2}c_2^r-16 e_{38}^r-2 e_{115}^r-2e_{116}^r 
\right)+O\left( M^5\right), 
\label{renmass1} 
\end{eqnarray} 
where ``r'' refers to $\widetilde{\rm MS}$-renormalized quantities. 
   When solving Eq.\ (\ref{nucleonmassdefinition}) in terms of 
Eqs.\ (\ref{sigmacontact}) and (\ref{olse}), we consistently omitted 
terms which count as $O(\hbar^2)$ in the loop expansion, i.e., terms 
proportional to $({\stackrel{\circ}{g_A}}_0/F_0)^4$, as well as terms 
proportional 
to $(c_1^r)^2$ which do not contribute in our final extended 
on-mass-shell expression for the nucleon mass. 
 
   Correspondingly, the wave function renormalization constant of the 
$\widetilde{\rm MS}$-renormalized propagator, 
\begin{displaymath} 
S_1(p)=\frac{1}{p\hspace{-.45em}/\hspace{.2em}-m 
-\Sigma_r(p\hspace{-.45em}/\hspace{.2em})} 
\to  \frac{Z_1}{p\hspace{-.45em}/\hspace{.2em}-m_N+i0^+}\,\, 
\mbox{for}\,\,p\hspace{-.45em}/\hspace{.2em}\to m_N, 
\end{displaymath} 
is an expression of ${\cal O}(q^3)$,\footnote{The reduction by one chiral 
order in comparison with the self-energy can be understood in 
terms of the derivative in the definition of the wave function renormalization 
constant.} 
given by 
\begin{equation} 
Z_1=\frac{1}{1-\Sigma_r'(m_N)}= 
1-\frac{9 {\stackrel{\circ}{g_{A}}}_r^2 M^2}{32 \pi^2 F^2_r}
\ln\left(\frac{M}{m}\right) 
-\frac{3 {\stackrel{\circ}{g_A}}_r^2 M^2}{16 \pi^2 F^2_r} 
+\frac{9 {\stackrel{\circ}{g_A}}_r^2 M^3}{64 \pi F^2_r m} 
\label{z1}. 
\end{equation} 
    Clearly, we do {\em not} require that the propagators of 
renormalized fields have unit residue at the physical pole mass 
\cite{Collins:xc,Itzykson:rh}. 
   The relation between $Z_1$ on the one hand and $Z_0$ and the field 
redefinition constant $\sqrt{Z_{01}}$ on the other hand is given by 
\begin{displaymath} 
Z_1=\frac{Z_0}{Z_{01}}. 
\end{displaymath} 
   Note, in particular, that $Z_1$ is finite, whereas both $Z_0$ and 
$Z_{01}$ contain infinities resulting from ultraviolet divergences. 
 
   In order to perform the second step, namely another {\em finite} 
renormalization, a given $\widetilde{\rm MS}$-renormalized diagram 
is written as the sum of a subtracted diagram which, through the 
application of the subtraction scheme described in the previous sections, 
satisfies the power counting and a remainder which violates the power 
counting and thus still needs to be subtracted. 
   We expand the finite renormalized couplings of the $\widetilde{\rm MS}$ 
scheme in a series in terms of couplings of our generalized 
on-mass-shell scheme. 
   In doing so, we generate finite counterterms, responsible for additional 
finite subtractions. 
   These counterterms are fixed so that the net result of combining 
the counterterm diagrams with those parts of the 
$\widetilde{\rm MS}$-renormalized diagrams which violate the power counting 
are of the same order as the subtracted diagram. 
(Note that depending on the applied renormalization condition the net result 
may vanish.) 
   Hence the sum of an $\widetilde{\rm MS}$-renormalized diagram and 
the corresponding counterterm diagram satisfies the power counting. 
 
   For the case at hand, we determine the terms to be subtracted from 
$\Sigma_a$ by first expanding the integrands and coefficients in 
Eq.\ (\ref{sigmaaresult}) in powers of $M^2$, 
$p\hspace{-.45em}/\hspace{.1em}-m$ and $p^2-m^2$. 
   In this expansion we keep all the terms having a chiral order which is 
smaller than what is suggested by the power counting for the given diagram. 
   We then obtain 
\begin{equation} 
\label{sesubtrterm} 
\Sigma_{r,a+b}^{\rm subtr}= \frac{3{\stackrel{\circ}{g_A}}_r^2}{32\pi^2 F_r^2} 
\left[mM^2-\frac{(p^2-m^2)^2}{4m}\right] 
+\frac{3 c_1^r {\stackrel{\circ}{g_A}}_r^2 M^2}{8 \pi^2 F_r^2} 
\left[m(p\hspace{-.45em}/\hspace{.1em}+m)-\frac{3}{2}(p^2-m^2)\right]. 
\end{equation} 
Equation (\ref{sesubtrterm}) specifies the part of the self-energy 
diagram which has to be subtracted. 
   We fix the corresponding counterterms so that they exactly cancel the 
expression given by Eq.\ (\ref{sesubtrterm}). 
   Since the most general Lagrangian contains all the structures consistent 
with the symmetries of the theory, it also provides the required counterterms. 
   Finally, the renormalized self-energy expression is obtained by subtracting 
Eq.\ (\ref{sesubtrterm}) from the  $\widetilde{\rm MS}$-subtracted version of 
Eqs.\ (\ref{sigmaa}) and (\ref{sigmab}) and replacing the 
$\widetilde{\rm MS}$-renormalized couplings with the ones of our generalized 
on-mass-shell scheme. 
   We note that the $\widetilde{\rm MS}$-subtracted version for $\Sigma_c$ 
needs no further subtraction because it is already of order ${\cal O}(q^4)$. 
 
    The correction to the nucleon mass resulting from the counterterms 
is calculated by substituting $p\hspace{-.45em}/\hspace{.1em}=m_N$ into 
the negative of Eq.\ (\ref{sesubtrterm}). 
[Recall that Eq.\ (\ref{sesubtrterm}) 
has to be subtracted.] 
    We thus obtain the following expression for the contribution to the mass: 
\begin{equation} 
\Delta m =-\frac{3 {\stackrel{\circ}{g_A}}^2 M^2}{32 \pi^2 F^2}(m+8c_1m^2). 
\label{deltam} 
\end{equation} 
   Finally, we express the physical mass of the nucleon up to and including 
order $q^4$ as \cite{Leutwyler:1999mz,Sainio:2001bq}\footnote{In our 
convention $k_3$ is larger by a factor of two than in Refs.\ 
\cite{Leutwyler:1999mz,Sainio:2001bq}, because we use $\ln(M/m)$ instead 
of $\ln(M^2/m^2)$.} 
\begin{equation} 
\label{massnucleonparameterization} 
m_N=m+k_1 M^2+k_2 M^3+k_3 M^4\ln\left(\frac{M}{m}\right)+k_4 M^4+ 
O(M^5), 
\end{equation} 
where $m$ is the nucleon mass in the chiral limit and $M^2=2 B \hat{m}$ is 
the leading-order result for $M_\pi^2$. 
  In terms of the EOMS-renormalized parameters, the coefficients $k_i$ 
are then given by 
\begin{eqnarray} 
\label{parki} 
k_1&=&-4 c_1\nonumber\\ 
k_2&=&-\frac{3 {\stackrel{\circ}{g_A}}^2}{32\pi F^2},\nonumber\\ 
k_3&=&\frac{3}{32\pi^2 F^2}\left(8c_1-c_2-4 c_3
-\frac{{\stackrel{\circ}{g_A}}^2}{m}\right), 
\nonumber\\ 
k_4&=&\frac{3 {\stackrel{\circ}{g_A}}^2}{32 \pi^2 F^2 m}(1+4 c_1 m) 
+\frac{3}{128\pi^2F^2}c_2-16 e_{38}-2 e_{115}-2e_{116}. 
\end{eqnarray} 
   Comparing with Ref.\ \cite{Becher:1999he}, we see that the lowest-order 
correction ($k_1$ term) and those terms which are nonanalytic in the quark 
mass $\hat{m}$ ($k_2$ and $k_3$ terms) coincide. 
   On the other hand, the analytic $k_4$ term ($\sim M^4$) is different. 
   This is not surprising, because we use a different renormalization scheme 
and hence the difference between the two results is compensated by different 
values of the renormalized parameters. 
 
   The contribution of the counterterms in Eq.\ (\ref{deltam}) to 
the physical mass is generated by the following expansion of the 
coupling $c_1^r$ in terms of our renormalized parameters: 
\begin{equation} 
c_1^r=c_1+\frac{3 m {\stackrel{\circ}{g_A}}^2}{128 \pi^2 F^2}\left[ 1+8 mc_1 
\right]+\cdots, 
\label{barecouplings} 
\end{equation} 
while the net result of the contributions of the counterterms which are 
generated by expanding the other parameters vanishes at the given 
order. 
   Finally, the wave function renormalization (residue at the pole) does not 
obtain a contribution from counterterms at this order so that $Z$ for 
our renormalization scheme reads 
\begin{equation} 
Z= 
1-\frac{9 {\stackrel{\circ}{g_A}}^2 M^2}{32 \pi^2 F^2}
\ln\left(\frac{M}{m}\right) 
-\frac{3 {\stackrel{\circ}{g_A}}^2 M^2}{16 \pi^2 F^2} 
+\frac{9 {\stackrel{\circ}{g_A}}^2 M^3}{64 \pi F^2 m}. 
\label{Z} 
\end{equation} 
 
\section{\label{conclusions}Conclusions} 
   We have discussed a new renormalization scheme which allows for a
simple and consistent power counting in the single-nucleon sector
of relativistic chiral perturbation theory.
   In order to renormalize a given diagram, using Eq.\ (\ref{dimension})
one first assigns a chiral order $D$ to that diagram.
   Applying standard techniques the diagram is reduced to the sum of 
dimensionally regularized scalar integrals multiplied by corresponding Dirac 
structures.
   By expanding the integrands as well as the coefficients in small 
quantities one identifies those terms which need to be subtracted in order
to produce the renormalized diagram with the chiral order $D$ determined 
beforehand.
   It is this aspect which we refer to as ``appropriately chosen 
renormalization conditions,'' because these subtractions can be realized in 
terms of local counterterms in the most general effective Lagrangian.
   For pedagogical reasons we have performed the subtractions in two
steps: the first step, namely, applying a modified minimal subtraction
scheme (of ChPT) to get rid of the ultraviolet divergences, corresponds to
the procedure used by Gasser, Sainio, and \v{S}varc \cite{Gasser:1988rb}.
   In a second step we have then performed additional {\em finite} 
subtractions for those integrals which contain nucleon propagators such
that the subtracted diagram satisfies our power counting scheme.    
   We have explicitly applied our scheme to a calculation of the nucleon
mass.
   Comparing with the results of the infrared regularization method 
\cite{Becher:1999he} we have seen that the expressions for the nucleon mass
in the two schemes only differ by terms which are analytic in the quark masses.
   These findings are consistent, because such terms are renormalization-scheme
dependent.

   Finally, our renormalization scheme is neither restricted to the 
single-nucleon sector nor to the interaction of Goldstone bosons with 
fermions.
   For example, it may also be used in the NN sector or for describing
the interaction of vector and axial-vector mesons \cite{Gegelia:2003b}.
   In conclusion, we have presented a simple renormalization scheme 
which produces a consistent power counting for relativistic baryon
chiral perturbation theory.
 
\acknowledgments 
   The work of T.F.~and S.S.~was supported by the Deutsche 
Forschungsgemeinschaft (SFB 443).
   J.G.~acknowledges the support of the Alexander von Humboldt Foundation. 
   G.J.~was supported by NSF grant OPP-0236449/G067771.
 
\appendix
\section{\label{appendix_counterterms}The generation of counterterms}
   The renormalization of the effective field theory (of pions 
and nucleons) is performed by expressing all the bare parameters and bare 
fields of the effective Lagrangian in terms of renormalized quantities 
\cite{Collins:xc}. 
   In this process, one generates counterterms which are responsible for the 
absorption of all the divergences occurring in the calculation of loop 
diagrams. 
   In order to illustrate the procedure let us discuss 
${\cal L}_{\pi N}^{(1)}$ 
and consider the free part in combination with the $\pi N$ 
interaction term with the smallest number of pion fields, 
\begin{equation} 
{\cal L}_{\pi N}^{(1)}=\bar \Psi_0 \left( i\gamma_\mu 
\partial^\mu -m_0 -\frac{1}{2}\frac{{\stackrel{\circ}{g_{A}}}_0}{F_0} 
\gamma_\mu 
\gamma_5 \tau^a \partial^\mu \pi^a_0\right) \Psi_0 +\cdots, 
\label{pieceoflolagr} 
\end{equation} 
given in terms of bare fields and parameters denoted by subscripts 0. 
   Introducing renormalized fields (we work in the isospin-symmetric 
limit) through 
\begin{equation} 
\label{renf} 
\Psi=\frac{\Psi_0}{\sqrt{Z_\Psi}},\quad 
\pi^a=\frac{\pi^a_0}{\sqrt{Z_\pi}}, 
\end{equation} 
we express the field redefinition constants $\sqrt{Z_\Psi}$ and 
$\sqrt{Z_\pi}$ 
and the bare quantities in terms of renormalized parameters: 
\begin{eqnarray} 
Z_\Psi&=& 1+\delta Z_\Psi\left(m,\stackrel{\circ}{g_{A}}, g_i, \nu \right), 
\nonumber\\ 
Z_{\pi}&=&1+\delta Z_\pi\left(m, \stackrel{\circ}{g_{A}}, g_i, \nu \right), 
\nonumber \\ 
m_0 &=& m(\nu )+\delta m\left( m, \stackrel{\circ}{g_{A}}, g_i, 
\nu \right), \nonumber\\ 
{\stackrel{\circ}{g_{A}}}_0&=&\stackrel{\circ}{g_{A}}(\nu)+\delta g_A \left( 
m, \stackrel{\circ}{g_{A}}, g_i, \nu \right), 
\label{bare} 
\end{eqnarray} 
where $g_i$, $i=1,\cdots \infty$, collectively denote all the renormalized 
parameters which correspond to bare parameters ${g_i}_0$ of the full effective 
Lagrangian. 
   The parameter $\nu$ 
indicates the dependence on the choice of the 
renormalization prescription.\footnote{Note that our choice 
$m(\nu 
)=m$, where $m$ is the nucleon pole mass in the chiral limit, is 
only one among an infinite number of possibilities.} 
   Substituting Eqs.\ (\ref{renf}) and (\ref{bare}) into 
Eq.\ (\ref{pieceoflolagr}), we obtain 
\begin{equation} 
\label{bct} 
{\cal L}_{\pi N}^{(1)}={\cal L}_{\rm basic}+{\cal L}_{\rm ct}+\cdots 
\end{equation} 
with the so-called basic and counterterm Lagrangians, respectively,\footnote{ 
Reference \cite{Collins:xc} uses a slightly different convention which 
is obtained through the replacement 
$(\delta Z_\Psi m+Z_\Psi\delta m)\to\delta m$.} 
\begin{eqnarray} 
\label{lbasic} 
{\cal L}_{\rm basic}&=&\bar \Psi \left( i\gamma_\mu 
\partial^\mu -m -\frac{1}{2} \frac{\stackrel{\circ}{g_{A}}}{F}\gamma_\mu 
\gamma_5 \tau^a \partial^\mu \pi^a\right) \Psi,\\ 
\label{lcounterterm} 
{\cal L}_{\rm ct}&=&\delta Z_\Psi \bar \Psi i\gamma_\mu\partial^\mu 
\Psi 
-\delta\{m\}\bar{\Psi}\Psi 
-\frac{1}{2}\delta\left\{\frac{\stackrel{\circ}{g_{A}}}{F}\right\} 
\bar{\Psi}\gamma_\mu \gamma_5 \tau^a \partial^\mu \pi^a \Psi, 
\end{eqnarray} 
where we introduced the abbreviations 
\begin{eqnarray*} 
\delta\{m\}&\equiv&\delta Z_\Psi m+Z_\Psi\delta m,\\ 
\delta\left\{\frac{\stackrel{\circ}{g_{A}}}{F}\right\}&\equiv& 
\delta Z_\Psi \frac{\stackrel{\circ}{g_{A}}}{F}\sqrt{Z_\pi} 
+Z_\Psi\left(\frac{{\stackrel{\circ}{g_{A}}}_0}{F_0}- 
\frac{\stackrel{\circ}{g_{A}}}{F} 
\right)\sqrt{Z_\pi} 
+\frac{\stackrel{\circ}{g_{A}}}{F}(\sqrt{Z_\pi}-1). 
\end{eqnarray*} 
   In Eq.\ (\ref{lbasic}), $m$, $\stackrel{\circ}{g_A}$, and $F$ denote 
the chiral limit of the physical nucleon mass, the axial-vector coupling 
constant, and the pion-decay constant, respectively. 
   Expanding the counterterm Lagrangian of Eq.\ (\ref{lcounterterm}) 
in powers of the renormalized coupling constants generates an infinite 
series, the individual terms of which are responsible for the subtraction 
of loop diagrams. 

\section{\label{loop_integrals}Loop integrals} 
   In this appendix we collect the loop integrals needed in the calculation 
of the nucleon mass. 
   Most of them can be found in Ref.\ \cite{Becher:1999he} or have 
been calculated using the method of dimensional counting 
\cite{Gegelia:1994zz}. 
   We use the following convention for scalar loop integrals: 
\begin{equation} 
\label{iuconvention} 
I_{N\cdots\pi\cdots}(p_1,\cdots,q_1,\cdots) 
=i\int\frac{d^n k}{(2\pi)^n}\frac{1}{[(k+p_1)^2-m^2+i0^+]\cdots 
[(k+q_1)^2-M^2+i0^+]\cdots}. 
\end{equation} 
   Tensor integrals are then derived in the standard fashion (see, 
for example, Appendix C of Ref.\ \cite{Scherer:2002tk}). 
   We do not display terms of $O(n-4)$ and higher. 
   In what follows, $\bar{\lambda}$ is defined as 
\begin{equation} 
\bar\lambda ={m^{n-4}\over 16\pi^2}\left\{ {1\over n-4}-{1\over 2} 
\left[ \ln (4\pi) +\Gamma '(1)+1\right]\right\}. 
\label{lambdabarapp} 
\end{equation} 
   From the set of purely pionic integrals we need  
\begin{equation} 
\label{Ipi} 
I_\pi(q)=I_\pi(0)=I_\pi=i\int \frac{d^n k}{(2\pi)^n}\frac{1}{k^2-M^2+i0^+} 
=2M^2\bar{\lambda}+\frac{M^2}{8\pi^2}\ln\left(\frac{M}{m}\right).
\end{equation} 
   The $\widetilde{\rm MS}$-renormalized integral is obtained by simply
dropping the term proportional to $\bar{\lambda}$:
\begin{equation} 
\label{Ipir} 
I_\pi^r=\frac{M^2}{8\pi^2}\ln\left(\frac{M}{m}\right). 
\end{equation} 
   Next we consider the integral containing only a nucleon propapagator: 
\begin{equation} 
\label{IN} 
I_{N}(p)=I_N(0)=I_N=i\int \frac{d^n k}{(2\pi)^n} \frac{1}{k^2-m^2+i0^+} 
=2m^2\bar{\lambda}.
\end{equation} 
   The $\widetilde{\rm MS}$-renormalized integral then reads
\begin{equation} 
\label{INr} 
I_N^r=0.
\end{equation} 
   Finally, we list the relevant integrals containing both a pion and
a nucleon propagator:
\begin{eqnarray} 
\label{INpi} 
I_{N\pi}(p,0)&=&i\int\frac{d^n k}{(2\pi)^n}\frac{1}{[(k+p)^2-m^2+i0^+] 
[k^2-M^2+i0^+]}\nonumber\\ 
&=&2\bar{\lambda}+\frac{1}{16\pi^2}\left[-1
+\frac{p^2-m^2+M^2}{p^2}\ln\left(\frac{M}{m}\right)
+\frac{2mM}{p^2}F(\Omega)\right],
\end{eqnarray}
where
\begin{eqnarray*} 
F(\Omega) 
&=& 
\left \{ \begin{array}{ll} 
\sqrt{\Omega^2-1}\ln\left(-\Omega-\sqrt{\Omega^2-1}\right),&\Omega\leq -1,\\
\sqrt{1-\Omega^2}\arccos(-\Omega),&-1\leq\Omega\leq 1,\\
\sqrt{\Omega^2-1}\ln\left(\Omega+\sqrt{\Omega^2-1}\right)
-i\pi\sqrt{\Omega^2-1},&1\leq
\Omega,
\end{array} \right. 
\end{eqnarray*} 
with 
\begin{displaymath} 
\Omega=\frac{p^2-m^2-M^2}{2mM}.
\end{displaymath} 
   Correspondingly,
\begin{equation}
\label{INpir}
I_{N\pi}^r(p,0)=I_{N\pi}(p,0)-2\bar{\lambda}.
\end{equation}
   Furthermore we need
\begin{eqnarray}
\label{Imunpi}
I^\mu_{N\pi}(-p,0)&=&i\int\frac{d^n k}{(2\pi)^n}
\frac{k^\mu}{[(k-p)^2-m^2+i0^+] [k^2-M^2+i0^+]}\nonumber\\ 
&=&\frac{p^\mu}{2p^2}\left[(p^2-m^2+M^2)I_{N\pi}(-p,0)+I_N
-I_\pi\right].
\end{eqnarray}

\end{document}